\documentclass[12pt]{article}
\usepackage{graphicx}
\usepackage{graphics}
\usepackage{amssymb}
\usepackage{amscd}
\begin{document}
%\begin{titlepage}

%\begin{center}
%{\hbox to\hsize{
%\hfill \bf hep-ph/??? }}

\bigskip \vspace{3\baselineskip}

\begin{center}
{\bf \large 
Cosmological matter-antimatter asymmetry \\ as a quantum fluctuation}

\bigskip

\bigskip

{\bf Archil Kobakhidze and Adrian Manning \\ }

\smallskip

{ \small \it
ARC Centre of Excellence for Particle Physics at the Terascale, \\
School of Physics, The University of Sydney, NSW 2006, Australia \\
E-mail: archilk@physics.usyd.edu.au, a.manning@physics.usyd.edu.au
\\}

\bigskip
 
\bigskip

\bigskip

{\large \bf Abstract}

\end{center}
\begin{quote}
\small{
We entertain a new paradigm according to which  the observed matter-antimatter asymmetry is generated as a large-scale quantum fluctuation over the baryon-symmetric state that occurred during the cosmic inflation.}
\end{quote}

%\end{titlepage}

\vspace{1cm}

%\paragraph{1.} 

\paragraph{1.}

The celebrated theory of Big Bang Nucleosynthesis (BBN), while correctly predicting abundances of light elements, 
assumes baryon asymmetry in the universe, measured in terms of the asymmetry parameter $\eta_B$:
\begin{equation}
\eta_B = \frac{(n_B-n_{\bar B})}{s}\approx 8.6\cdot 10^{-11}~, 
\label{1}
\end{equation}
where $n_{B (\bar B)}$ is a baryon (antibaryon) number density and $s$ is an entropy density. Theoretically, this asymmetry is puzzling. Indeed, on a rather general ground one would expect equal number of photons, baryons and anti-baryons in the early radiation-dominated Universe. The baryons and anti-baryons would quickly annihilate, leaving essentially only photons at the BBN epoch (the annihilation catastrophe). Also, the asymmetry viewed as a statistical fluctuation over the baryon symmetric universe is $\eta_{B}\sim 1/\sqrt{n_{\gamma}}\sim 10^{-40}$, too small compared to the observed value (\ref{1}).  This picture is affected if non-equilibrium baryon-number and CP-violating interactions were active at some stage of the evolution of the universe, leading to a dynamical generation of the asymmetry \cite{Sakharov:1967dj}.  These conditions can be met in many particle physics models, including the minimal Standard Model. However, to generate the desired asymmetry is by far a trivial problem, and requires quite involved extensions of the Standard Model (see, e.g., \cite{Riotto:1999yt} for a review and references therein). 
 
Perhaps the failure to generate sizeable matter-antimatter asymmetry in a variety of particle physics models, particularly within the Standard Model, simply indicates  that the total number densities of baryons and anti-baryons are, in fact, equal in the whole universe, i.e. $n_B-n_{\bar B}\approx 0$. The asymmetry parameter $\eta_B$ can still be non-zero due to the non-zero  variance $\delta n_B\neq 0$ within the observable universe $\eta_{B}=\delta n_B/s$. Such a large-scale quantum fluctuation may have been produced during the cosmic inflation, without Sakharov's conditions \cite{Sakharov:1967dj} being satisfied\footnote{Recent models of inflationary baryogenesis within the standard paradigm can be found in \cite{Barrie:2014waa}. See also references therein.}. In what follows we demonstrate this idea within a toy model of a quantum scalar field carrying baryonic charge in the inflationary universe.         

\paragraph{2.} Let $\hat \phi(x)$ be  a complex scalar field carrying a unit U(1) (baryonic) charge,  which is strictly conserved. The conserved comoving charge operator can be conveniently expressed through the comoving field operator $\hat \Phi= \frac{1}{\sqrt{2}} \left(\hat \Phi_1+i\hat \Phi_2 \right)=a(\eta)\hat \phi$ as
\begin{equation}
\hat Q(\eta, \vec x)= \left(\hat \Phi_2\partial_{\eta}\hat \Phi_1-\hat \Phi_1\partial_{\eta}\hat \Phi_2\right)~,
\label{2}
\end{equation}
where $a(\eta)$ is a scale factor of a spatially flat inflationary universe and $\partial_{\eta}$ is a partial derivative with respect to conformal time $\eta$. We would like to evaluate the charge-charge correlator at the end of inflation over the comoving scale $\ell=|\vec x-\vec y|$:
\begin{eqnarray}
 \bar Q^2_{\ell}= \int d^3\vec x d^3\vec y  W_{\ell}(\vec x)W_{\ell}(\vec y)\langle \hat Q(\eta_{\rm inf}, \vec x) \hat Q(\eta_{\rm inf}, \vec y)\rangle~,
\label{3}
\end{eqnarray} 
where $W_{\ell}(\vec x)$ is taken as a Gaussian window function
\begin{eqnarray}
	W_{\ell}(\vec x) =\frac{1}{(2\pi)^{\frac{3}{2}} \ell^3} {\rm e}^{-|\vec x|^2/2\ell^2}~,
\label{4}
\end{eqnarray} 
We work in the de Sitter approximation for inflation, hence, $\eta_{\rm inf}=1/a_{\rm inf}H_{\rm inf}$, where $H_{\rm inf}$ and $a_{\rm inf}$ are the Hubble rate and scale factor at the end of inflation. The correlator $\langle \hat Q(x) \hat Q(x')\rangle$ in (\ref{3}) can be expressed in terms of the scalar two-point function,
\begin{eqnarray}
G(x,x')=\langle\hat \Phi_1(x) \hat \Phi_1(x')\rangle = 
\langle\hat \Phi_2(x) \hat \Phi_2(x')\rangle=
\int \frac{d^3\vec k}{(2 \pi)^3} \Phi_k (\eta) \Phi_k^* (\eta') {\rm e}^{i \vec{k}\cdot \left(\vec x- \vec{x'}\right)}  ~,\nonumber \\
\label{5}
\end{eqnarray}
as \cite{Calzetta:1997ku}:
\begin{eqnarray}
\langle \hat Q(x) \hat Q(x')\rangle= 2 \left[ G(x,x') \frac{\partial^2}{\partial \eta \partial \eta'} G(x,x') - \frac{\partial}{\partial \eta'} G(x,x') \frac{\partial}{\partial \eta} G(x,x') \right ]~.
\label{6}
\end{eqnarray}
In Eq. (\ref{5}),  
 \begin{eqnarray}
 \Phi_k(\eta) &= i\sqrt{-\frac{\pi \eta}{4}} H_{3/2}^{(1)} (-k \eta) \approx 
 \frac{1}{\sqrt{2 k}} \frac{1}{k|\eta|}\left[ 1 +\frac{(k\eta)^2}{2} \right] 
\label{7}
\end{eqnarray} 
are the de Sitter mode functions defined using the standard Bunch-Davis prescription for the vacuum state  \cite{Bunch:1978yq}. As we are primarily interested in large scale correlations, in the last equation of (\ref{7}) we have taken $k|\eta|\to 0$ limit and retain only terms up to the order $k|\eta|$. We also assume that mass of the scalar field is negligible compared to the inflationary rate.  

Using (\ref{7}), (\ref{5}), (\ref{6}) we compute the charge-charge correlator at the end of inflation (\ref{3}) and define the quantum fluctuation in the difference of particle and antiparticle number densities over the particle-antiparticle symmetric state as:
 \begin{eqnarray}
|\delta n_B|\equiv \bar Q_{\ell} \approx \frac{\sqrt{\kappa}}{4 \sqrt{2} \pi^2} \frac{a_{\rm \inf}H_{\rm inf}}{ \ell^2}~, 
\label{8}
\end{eqnarray} 
where $\kappa \approx \frac{1}{{\rm e}^2}\left[ 2{\rm e}{\rm Ei}(1)-1\right]\approx 0.026$ is a numerical factor resulting from integration.  We note that the above non-zero asymmetry first emerges in the ${\cal O}(k|\eta|)$ approximation for the mode functions (\ref{7}), the leading term ${\cal O}(1/k|\eta|)$ alone gives zero asymmetry.
Recall, (\ref{8}) is a comoving asymmetry in particle and antiparticle number densities homogeneously distributed within the comoving volume $\sim\ell^3$. To compute the matter-antimatter asymmetry parameter $\eta_B$ we divide (\ref{8}) by the comoving entropy density generated at the end of the reheating process. We obtain: 
   \begin{eqnarray}
|\eta_B|=\frac{|\delta n_B|}{a_{\rm rh}^3s}= 
\frac{45\sqrt{\kappa}}{8 \sqrt{2} \pi^4g_{*}(T_{\rm rh})}\left(\frac{a_{\rm inf}}{a_{\rm rh}}\right)^3 
\frac{1}{\left(a_{\rm inf}\ell\right)^2}
\frac{H_{\rm inf}}{ T^3_{\rm rh}}~,
\label{9}
\end{eqnarray} 
 where $a_{\rm rh}$ and $T_{\rm rh}$ are the scale factor and temperature at the end of reheating, while $g_{*}(T_{\rm rh})$ is a number of particle species produced during the reheating; typically $g_{*}(T_{\rm rh})\approx 100$ in the Standard Model. 
 
If our proposal is correct and assuming no further production of entropy\footnote{This approximation potentially breaks down when the Hubble radius of the universe becomes of the order of the scalar mass. The scalar condensate may start decaying into lighter particles, including ordinary quarks and anti-quarks. However, if the interactions with other fields are baryon-number conserving, it is reasonable to expect that the net asymmetry will not be affected significantly and will be simply transferred into ordinary matter. Sphaleron processes at the electroweak scale may also contribute to reprocessing the generated baryon number into a lepton number, however, this cannot change baryon asymmetry significantly because of B-L conservation.}, the baryon asymmetry (\ref{9}) has to match with the  observed values at experiments (\ref{1}). The largest scales at which the imbalance between matter and antimatter is observed can be deduced from non-observation of distortion of CMB radiation. It has been demonstrated in \cite{Cohen:1997ac} that matter has to dominate over antimatter within the entire visible universe, i.e. at scales $L_0\approx 1.6\cdot 10^{41}$ 1/GeV. The corresponding comoving scale at the end of inflation is then
\begin{equation}
\ell = \left(\frac{a_{\rm rh}}{a_0}\right)\left(\frac{a_{\rm inf}}{a_{\rm rh}}\right)^{1/2}\frac{L_0}{a_0}~,
\label{10}
\end{equation}   
where $a_0$ denotes the present scale factor and we assume for simplicity a radiation dominated era from the end of reheating until today and approximate the reheating period with a matter-dominated era. Plugging (\ref{10}) into (\ref{9}) and using $a_0 / a_{\rm  rh}=T_{\rm rh}/T_0$ [$T_0\approx 10^{-13}$ GeV is the current CMB temperature] we obtain:
    \begin{eqnarray}
|\eta_B|= 
\frac{45\sqrt{\kappa}}{8 \sqrt{2} \pi^4g_{*}(T_{\rm rh})}
\frac{H_{\rm inf}}{L_0^2}\frac{T_{\rm rh}}{T_0^4}\approx 10^{-10}\left(\frac{H_{\rm inf}}{2\cdot 10^{11}~ {\rm GeV}}\right)\left(\frac{T_{\rm rh}}{2\cdot 10^{11}~ {\rm GeV}}\right)~.
\label{11}
\end{eqnarray} 
Thus, the observed asymmetry can indeed be accounted for by a mechanism described here for a large and realistic range of inflationary rates and reheating temperatures.

We stress again, that the asymmetry (\ref{11}) produced as a quantum fluctuation within a comoving particle horizon (of size $\ell$) at the end of inflation will stay essentially intact at any subsequent epochs, providing a necessary feed for a successful generation of light elements during the Big Bang Nucleosynthesis (BBN) era as well as fitting in the baryonic content of the universe as measured through the Cosmic Microwave Background radiation.  Note, that fluctuations at smaller subhorizon ($<\ell$ ) scales would result in domains with baryon and antibaryon dominance. Because the average baryon number in our model is equal to zero and, hence, the number of baryon and antibaryon domains are approximately equal, these baryon and antibaryon domains are expected to quickly annihilate leaving only the asymmetry (\ref{11}). Therefore, we expect that BBN and CMB eras will proceed in the standard way, unlike the models with separated baryon and antibaryon domains (see, e.g., a review \cite{Dolgov:1991fr} and references therein). The fact that the baryon number is globally zero, is also the key difference of our model from all the models based on Sakharov's scenario of baryogenesis, including the Affleck-Dine mechanism \cite{Affleck:1984fy}, where a non-zero baryon charge is stored in the scalar field condensate.        

\paragraph{3.} In this letter we have proposed a qualitatively new explanation for the matter-antimatter imbalance in the universe, which does not requires Sakharov's conditions to be satisfied. The idea is that the universe is fundamentally matter-antimatter symmetric, and the observed asymmetry represents a quantum fluctuation that occurred during the phase of cosmic inflation at a scale that corresponds to the size of visible universe today. We have performed explicit calculations within the simplified toy model and confirmed that sufficient asymmetry can indeed be generated. We believe this toy model can be extended to more realistic theories. Furthermore, a similar mechanism can be used to generate asymmetric dark matter during inflation, e.g., along the lines discussed in Ref. \cite{Barrie:2015axa}.

\paragraph{Acknowledgements.} This work was partially supported by the Australian Research Council. AK was also supported in part by the Rustaveli National Science Foundation under the projects No. DI/8/6-100/12 and No. DI/12/6-200/13. 

%\newpage

\end{document}